\documentclass[debug,overfull]{epl}
\usepackage{graphicx}% Include figure files
\usepackage{dcolumn}% Align table columns on decimal point
\usepackage{bm}% bold math
\usepackage{float}

\title{Statistical Properties of Business Firms Structure and Growth }

\author{Kaushik Matia\inst{1}, Dongfeng Fu\inst{1},
Sergey~V.~Buldyrev\inst{1}, Fabio Pammolli\inst{2}, Massimo
Riccaboni\inst{2} and H. Eugene~Stanley\inst{1}}

\institute{
\inst{1}Center for Polymer Studies and
Department of Physics, Boston University, Boston, MA 02215 USA\\ 
\inst{2}Faculty of Economics, University of Florence and CERM, 
Via Banchi di Sotto 55, 53100 Siena Italy}
\begin{document}
\maketitle
\begin{abstract}

We analyze a database comprising quarterly sales of 55624 pharmaceutical
products commercialized by 3939 pharmaceutical firms in the period
1992--2001. We study the probability density function (PDF) of growth in
firms and product sales and find that the width of the PDF of growth
decays with the sales as a power law with exponent $\beta = 0.20 \pm
0.01$. We also find that the average sales of products scales with the
firm sales as a power law with exponent $\alpha = 0.57 \pm 0.02$. And
that the average number products of a firm scales with the firm sales as
a power law with exponent $\gamma = 0.42 \pm 0.02$. We compare these
findings with the predictions of models proposed till date on growth of
business firms.
\end{abstract}

In economics there are unsolved problems that involve interactions among
a large number of subunits~\cite{Ijiri97,Sutton02,Wyart}. One of these
problems is the structure of a business firm and its
growth~\cite{Sutton02,Coase}. As in many physical models, decomposition
of a firm into its constituent parts is an appropriate starting place
for constructing a model. Indeed, the total sales of a firm is comprised
of a large number of product sales. Previously accurate data on the
``microscopic'' product sales have been unavailable, and hence it has
been impossible to test the predictions of various models. Here we
analyze a new database, the Pharmaceutical Industry Database (PHID)
which records quarterly sales figures of 55624 pharmaceutical products
commercialized by 3939 firms in the European Union and North America
from September 1991 to June 2001. We shall see that these data support
the predictions of a simple model, and at the same time the data do not
support the microscopic assumptions of that model. In this sense, the
model has the same status as many statistical physics models, in that
the predictions can be in accord with data even though the details of
microscopic interactions are not. The assumptions of this simple model
given in Ref~\cite{PRL98} are as follows: (i) Firms tends to organize
itself into multiple divisions once they attain a particular size. (ii)
The minimum size of firms in a particular economy comes from a broad
distribution. (iii) Growth rates of divisions are independent of each
other and there is no temporal correlation in their growth. With
these assumptions the model builds a diversified multi-divisional
structure. Starting from a single product evolving to a multi-product
firm, this model reproduces a number of empirical observations and make
some predictions which we discuss in detail below along with results and
predictions of other models which attempt to address the problem of
business firm growth.

Consider a firm $i$ of sales $S_i$ with $N_i$ products whose sales are
$\xi_{i,j}$ where $j=1,2,...,N_i$. Thus the firm size in terms of the
sales of its products is given as $S_i = \sum_{j=1}^{j=N_i}
\xi_{i,j}$. The growth rate is
\begin{equation}
g_i(t) \equiv \log \left(\frac{S_i(t+\Delta t)}{S_i(t)} \right) = \log S_i(t+\Delta t) - \log S_i(t),
\label{e.grate}
\end{equation}
where $S_i(t)$ and $S_i(t+\Delta t)$ are the sales, in units of $10^3$
British Pounds, of firm $i$ being considered in the year $t$ and
$t+\Delta t$, respectively. Pharmaceutical data has seasonal effect,
and hence the analysis of quarterly data will have effects due to
seasonality. To remove any seasonality, that might be present, we
analyze the annual data instead of the quarterly data.

Recent studies have demonstrated power-law scaling in economic systems
\cite{MStanley96}. In particular the standard deviation $\sigma$ of the
growth rates of diverse systems including firm sales ~\cite{MStanley96}
or gross domestic product (GDP) of countries~\cite{Plerou1} scales as a
power-law of $S$.

\begin{figure}
%\narrowtext
\begin{center}
  \includegraphics[width=0.38\textwidth, angle = -90]{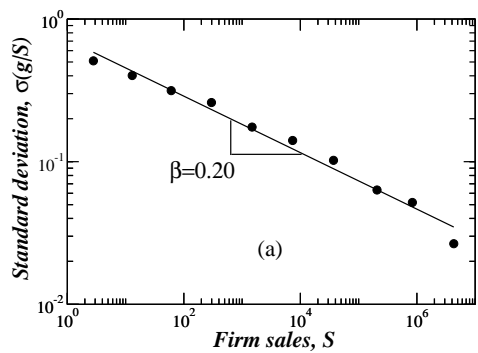}
\end{center}
\begin{center}
  \includegraphics[width=0.38\textwidth, angle = -90]{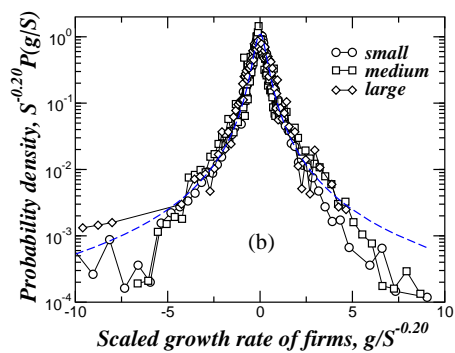}
\end{center}
\caption{(a) Firms are divided into 10 groups according to firm sale
$S$. We find the standard deviation $\sigma(g|S)$ of the growth rates
scales as a power law, $\sigma(g|S) \sim S^{-\beta} $ with $\beta=0.20
\pm 0.01$. Symbols are data points and solid line is a regression
fit. (b) PDF of the growth rates for small [$S< 10^2$], medium [$10^2 <
S < 10^4$], and large [$10^4 < S $] values of S is scaled by their
standard deviation. Note the collapse of the histograms of the three
groups which confirms the scaling exponent $\beta$. The dashed line is
$f_0$ as predicted by the model approximating the results of ref
{~\protect\cite {PRL98}} given by $f_0(x) \approx a_0 \exp \{
(-a_1~(1+0.75~\ln (1+a_2~ x^2)]\}[1+0.75~\ln(1+a_2~x^2)]^{1/2}$ where
$a_0,~a_1$,and $a_2$ are parameters of the model.}
\label{Fig1}
\end{figure}

\begin{figure}
%\narrowtext
%\vspace*{0.6cm}
%\centerline{
\begin{center}
\includegraphics[width=0.38\textwidth,angle=-90]{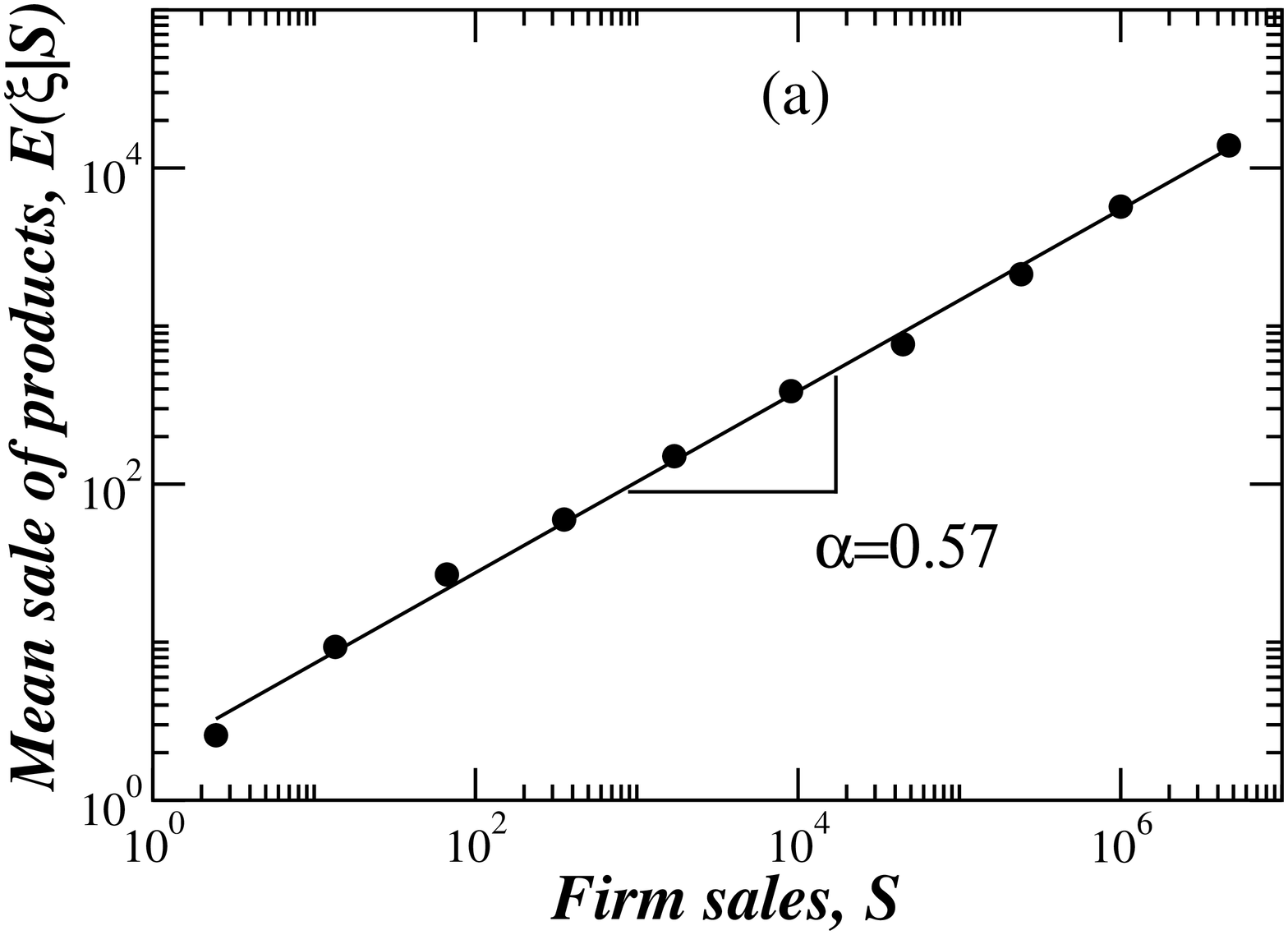}
\end{center}
%}
\begin{center}
\includegraphics[width=0.38\textwidth,angle=-90]{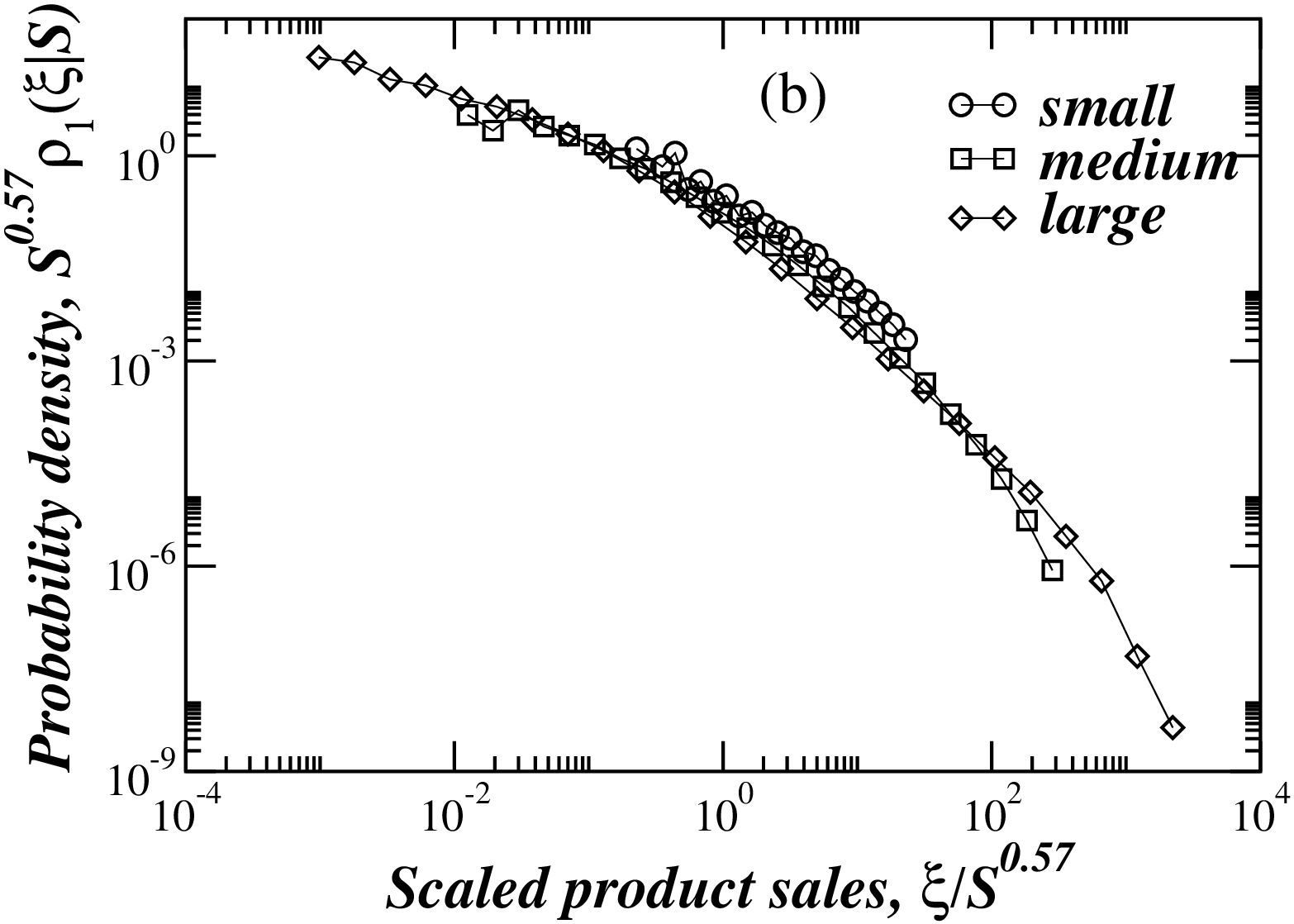}
%\vspace*{0.6cm}
\end{center}
\caption{(a) Mean, $E(\xi|S)$ of the product sale conditioned for a firm
of sale $S$. We observe that the mean scales as $E(\xi|S) \sim
S^{\alpha}$ with $\alpha = 0.57 \pm 0.02$. Symbols are data points and
solid line is a regression fit. (b) PDF of the product sales for small
[$S< 10^2$], medium [$10^2 < S < 10^4$], and large [$10^4 < S $] values
of S scaled by $S^{0.57}$. Note the collapse of the PDF's of the three
groups which confirms the scaling exponent $\alpha$. }
\label{Fig2}
\end{figure}

The models of Refs~\cite{Ijiri97,PRL98,Fabritiis,Sutton02,Wyart} all
predicts that standard deviation of growth rates amongst all firms with
the same sales scales as a power law $\sigma(g|S) \sim
S^{-\beta}$. Further, model of Refs~\cite{PRL98} predicts that
probability density function PDF $p(g | S)$, of growth rates for a size
of firm $S$ scales as a function of $S$ as :
\begin{equation}
p(g | S) \sim \frac{1}{S^{-\beta}} ~~ f_0\left( \frac{g}{S^{-\beta}} \right)\,\,
\label{e.1}
\end{equation}
where $f_0$ is a symmetric function of a specific ``tent-shaped'' form
resulting from a convolution a log normal distributions and a Gaussian
distribution, with parameters dependent on the parameters of the
model. Figure~\ref{Fig1}a plots the scaling of the standard deviation
$\sigma(g| S)$. We observe $\sigma(g|S) \sim S^{-\beta} $ with $\beta =
0.19 \pm 0.01$. Figure~\protect\ref{Fig1}b plots the scaled PDF as given
by eq.~\ref{e.1} for three sales groups; small ($ S <10^2$), medium ($
10^2<S <10^4$) and large ($10^4<S$). The figure also plots $f_0$ as
predicted by refs~\cite{PRL98}.
The model of Ref~\cite{PRL98} further predicts that the PDF of the
product size $\xi$ for a fixed firm size $S$, $\rho_1(\xi | S)$ should
scale as
\begin{equation}
\rho_1(\xi | S) \sim \frac{1}{S^{\alpha}} ~~ f_1\left( \frac{\xi}{S^{\alpha}} \right)\,.
\label{e.2}
\end{equation}
where again $f_1$ depends on the parameters of the model. According to
the model discussed in Refs~\cite{PRL98,Fabritiis} $f_1$ is a log-normal
PDF. We then evaluate the average product size $E(\xi|S)$ in a firm of
size $S$, defined as $E(\xi | S) = \int d\xi \rho_1(\xi | S) \xi \sim
S^{\alpha}$. Figure~\ref{Fig2}b plots $E(\xi|S)$, we observe $E(\xi|S)
\sim S^{\alpha}$ with $\alpha = 0.57 \pm
0.02$. Figure~\protect\ref{Fig2}b plots the scaled PDF as given by
eq.~\ref{e.2} for three sales groups; small ($ S <10^2$), medium ($
10^2<S <10^4$) and large ($10^4<S$). We observe that for each of the
groups the PDF $\rho_1(\xi| S)$ is consistent with a log-normal
distribution by noting in a log-log plot the PDF $\rho_1(\xi| S)$ is
parabolic which is tested by performing a regression fit.
\begin{figure}
%\narrowtext
%\vspace*{0.6cm}
%\centerline{
\begin{center}
\includegraphics[width=0.38\textwidth,angle=-90]{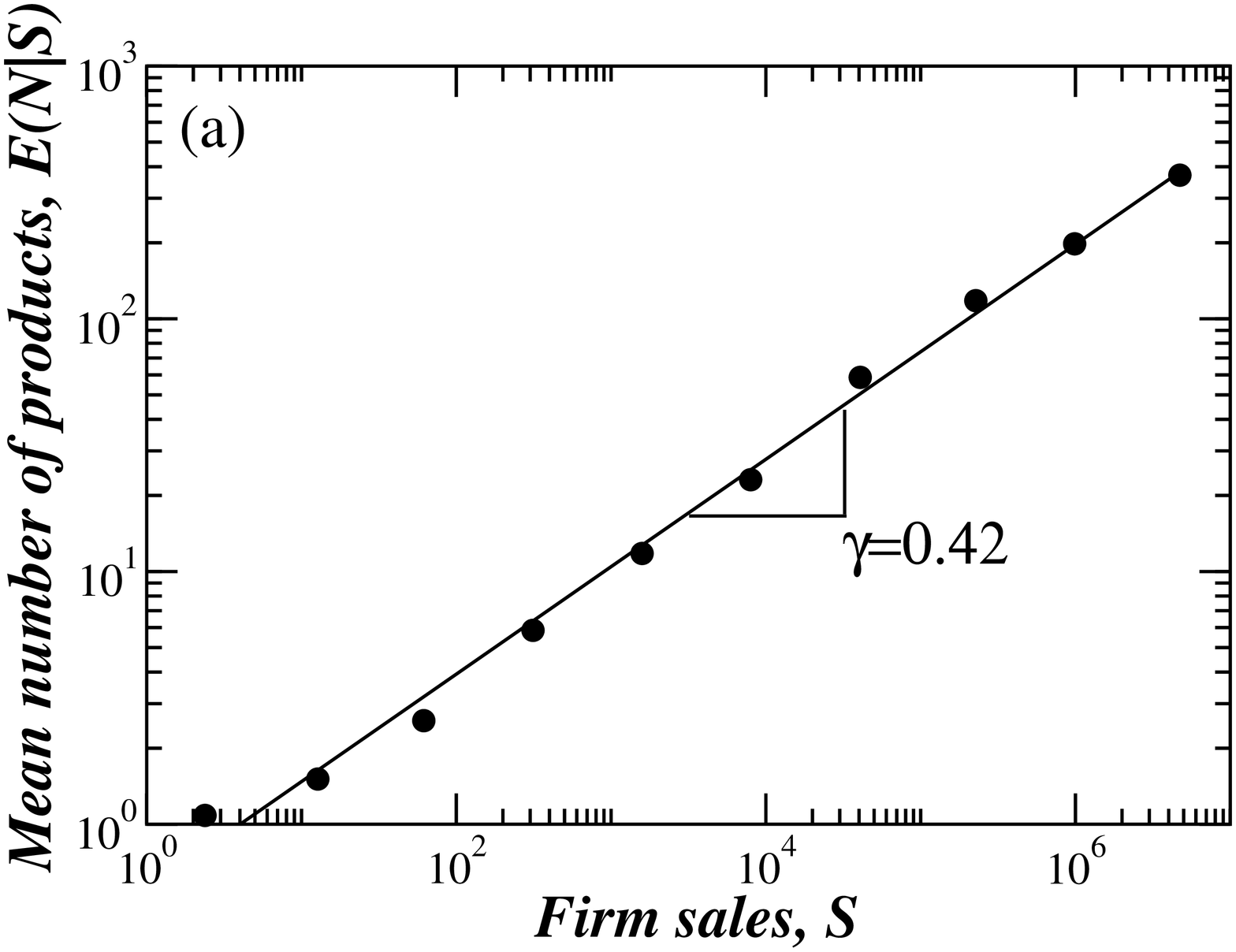}
\end{center}
\begin{center}
\includegraphics[width=0.38\textwidth,angle=-90]{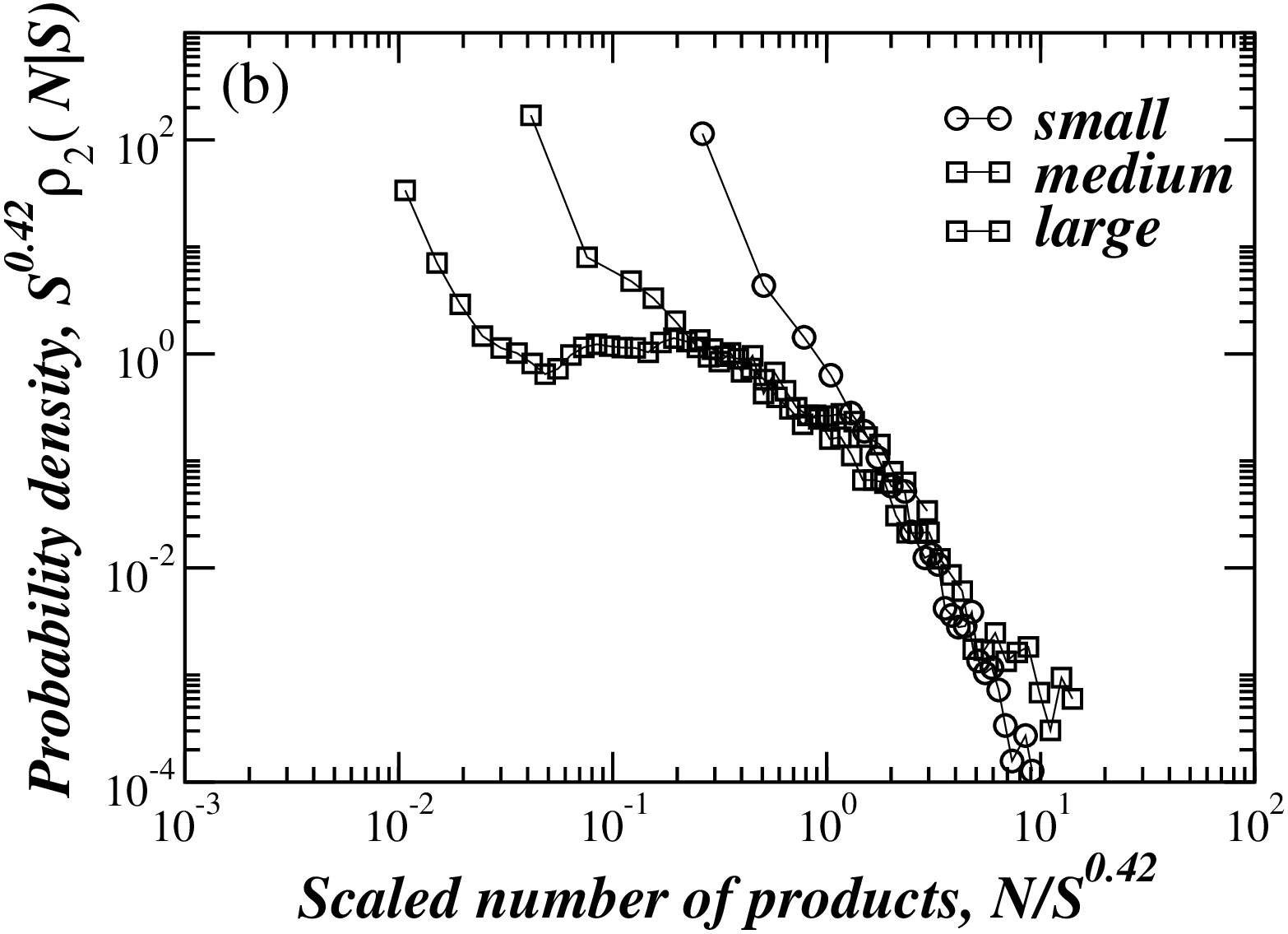}
\end{center}
%\vspace*{0.6cm}
\caption{(a) Mean, $E(N|S)$ of the number of products $N$ for a firm of
sale $S$. Symbols are data points and solid line is a regression fit. We
observe that the mean scales as $E(N|S) \sim S^{\gamma}$ with $\gamma =
0.42\pm 0.01$.(b) PDF of the number of product for small [$S< 10^2$],
medium [$10^2 < S < 10^4$], and large [$10^4 < S $] values of S scaled
by $S^{0.42}$. Note the partial collapse of the PDF's for of the three
groups which confirms the scaling exponent $\gamma$. For small values of
$N$, which also corresponds to small values of $S$, the statistics become
poor. This statistical errors gets even more amplified when we divide
small values of $N$ by $S^{0.42}$. Thus we observe poor quality of data
collapse for $N/S^{0.42} < 1$. The data collapse is better for $N/S^{0.42}
> 1$ where we have good statistics. }
\label{Fig3}
\end{figure}
\begin{figure}
%\narrowtext
%\vspace*{0.6cm}
%\centerline{
\begin{center}
\includegraphics[width=0.38\textwidth,angle=-90]{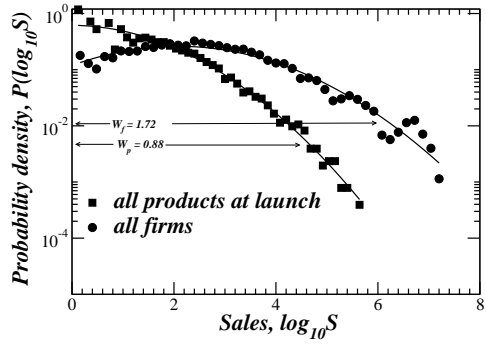}
%}
\end{center}
%\vspace*{0.6cm}
\caption{ PDF of products sales (diamond) firm sales (circles) between
1990-2001. The variance of the PDF's of products sales at launch and
firm sales are estimated to be $W_p =0.88$ and $W_f=1.72$
respectively. This gives $W_f -W_p=0.84$ and $\beta = 0.24$
[cf. eq.~\protect{\ref{e.5}}] which is approximately what is observed
empirically as predicted by~\protect{\cite{PRL98}}.}
\label{Fig4}
\end{figure}
\begin{figure}
%\narrowtext
\centerline{
%\begin{center}
\includegraphics[width=0.38\textwidth, angle=-90]{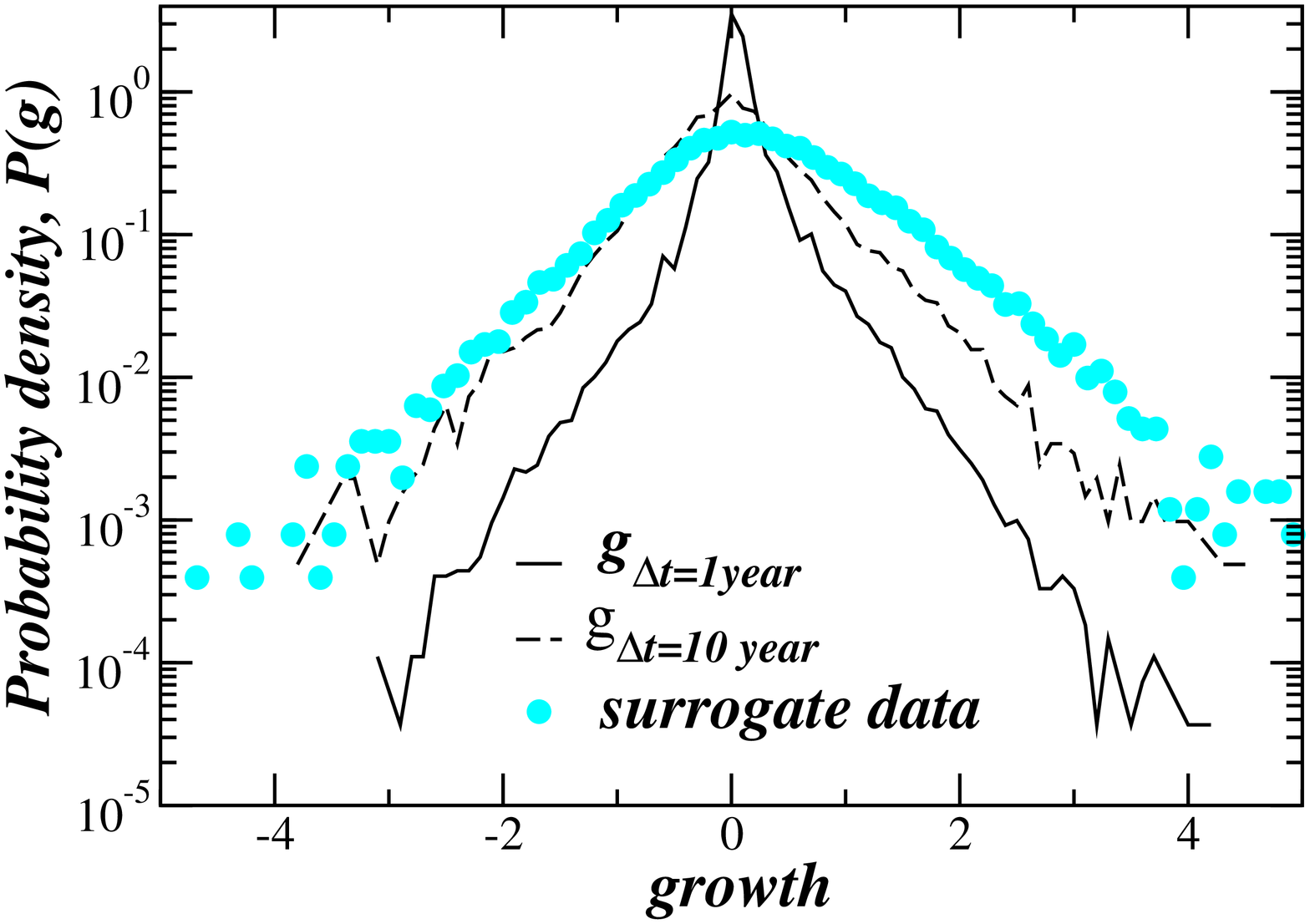}
%\end{center}
}
%\vspace*{0.1cm}
\caption{ Probability density function (PDF), $P(g_{\Delta t})$ of
products for ${\Delta t} = 1$ year (solid line) and ${\Delta t} = 10$
years (dashed line). Circles represent the of surrogate
data~\protect{\cite{Note1}}. In absence of correlation we expect the
data for $\Delta t=10$ to coincide with the PDF of the surrogate
points. }
\label{Fig5}
\end{figure}
According to Ref~\cite{PRL98}, the PDF $\rho_2(N | S)$ of number of
products $N$ in a firm of size $S$ should obey the scaling relation:
\begin{equation}
\rho_2(N | S) \sim \frac{1}{S^{\gamma}} ~~ f_2\left( \frac{N}{S^{\gamma}} \right)\,.
\label{e.3}
\end{equation}
where the function $f_2$ is log-normal and depends of the parameters of
the model. We evaluate the average number of
products $E(N|S)$ for a firm of size $S$. Using eq.~\ref{e.3} we note
that $E(N | S) = \int dN \rho_1(N | S) N \sim
S^{\gamma}$. Figure~\ref{Fig3}a plots the expectation $E(N|S)$ and we
observe that $E(N|S) \sim S^{\gamma}$ with $\gamma = 0.42 \pm
0.01$. Figure~\ref{Fig3}b plots the scaled PDF $\rho_2(N | S)$ as given
by eq.~\ref{e.3} for three groups; small ($ S <10^2$), medium ($ 10^2<S
<10^4$) and large ($10^4<S$) .

According to~\cite{PRL98} the relations between the scaling exponents
$\alpha~\beta$, and $~\gamma$ are given by
\begin{eqnarray}
\gamma &= &1 - \alpha,  \\
\beta &= &\frac{1 - \alpha}{2}
\label{e.4}
\end{eqnarray}
which we find to be approximately valid for the PHID database.

According to the model discussed in refs~\cite{PRL98}, the distribution
of product sizes in each firm scales with the product size at launch
$S_{\mathrm{at~launch}}$, according to,
$\rho_0(\xi/S_{\mathrm{at~launch}})$ which is approximately
log-normal. Model~\cite{PRL98} postulates that the PDF of $S_{\mathrm{at
launch}}$ is log normal, i.e., $P(\log S_{\mathrm{at~launch}})$ is
Gaussian with variance $W_p$ and each firm is characterized by a fixed
value of $S_{\mathrm{at~launch}}$. Furthermore, ref.~\cite{PRL98}
predicts that the distribution of firm sales is close to log normal
i.e., the PDF $P(\log S)$ is Gaussian with variance $W_f$. With these
hypothesis ref.~\cite{PRL98} derives that,
\begin{equation}
\beta = \frac{W_f -W_p}{2~W_f}.
\label{e.5}
\end{equation}
Figure~\ref{Fig4} plots PDF of annual products sales, products sales at
launch $P(\log S_{\mathrm{at~launch}})$, and firm sales $P(\log S)$
between 1990-2001. The variance of the PDF's of products sales at launch
and firm sales are estimated to be $W_p =0.88$, $W_f=1.72$
respectively. This gives $\beta = (W_f-W_p)/2~W_f = 0.24$ which is
approximately what is observed empirically. We employ two methods to
estimate $W$ (the standard deviation) : (i) Estimate $W$ from the
definition, i.e. $W^2 = (1/(N-1))\sum_{i=1}^{i=N} (x_i-<x>)^2$ where
$\{x_1,x_2,...x_N\}$ is a set of data and $<x>$ is the mean of the set
$\{x_i\}$. (ii) First estimate the PDF from the set $\{x_i\}$, then
perform a regression fit with a log-normal function to the PDF. The
standard deviation $W$ will be one of the fitting parameter. Hence
estimate $W$ from the estimated parameter value from the least square
log-normal fit to the PDF. We observe both this method gives similar
values of $W$ and the ratio $\beta$ (cf. eq.\ref{e.5} ) remains
unchanged as long as we consistently use one of the 2 methods described
above. Our estimate of $W$ presented here is using the former method.

Ref.~\cite{PRL98} postulates products growth rate to be Gaussian and
temporally uncorrelated. To test this postulate figure~\ref{Fig5} plots
the PDF $P(g)$ of the growth $g$ of the products $\Delta t = 1 $ and
$\Delta t = 10 $ year~\cite{Note1}. We see that the empirical
distribution is not growing via random multiplicative process as
ref.~\cite{PRL98} postulates but has the same tent shape distribution as
the distribution of firm sales growth rate, suggesting that the products
themselves may not be elementary units but maybe comprised of smaller
interacting subunits. Figure~\ref{Fig5} also plots PDF $P(g^{\prime})$
surrogate data obtained by summation of the 10 annual growth rates from
the empirical distribution. We observe that the $P(g_{\Delta t = 10})$
for products differs from the surrogate data implying there are
significant anti-correlation in the growth dynamics between successive
years.

In summary we study the statistical properties of the internal structure
of a firm and its growth. We identify three scaling exponents relating
the (i) sales of the products, $\xi$ (ii) the number of products, $N$
and (iii) the standard deviation of the growth rates, $\sigma$, of a
firm with its sales $S$. Our analysis confirms the features predicted in
ref~\cite{PRL98}. However we find that the postulate of the model
namely: the growth rate of the products is uncorrelated and Gaussian is
not accurate. Thus the model of ref.~\cite{PRL98} can be regarded as a
first step towards the explanation of the dynamics of the firm growth.

%%%%%%%%%%%%%%%%%%%%%%%%%%%%%%%%%%%%%%%%%%%%%%%%%%%%%%%%
%%%%%%%%%%%%%%%%%%%%%%%%%%%%%%%%%%%%%%% ACKNOWLEDGMENTS
%%%%%%%%%%%%%%%%%%%%%%%%%%%%%%%%%%%%%%%%%%%%%%%%%%%%%%%%%%

We thank L. A. N. Amaral, S.~Havlin for helpful discussions and
suggestions and NSF and Merck Foundation (EPRIS Program) for financial
support.
%%%%%%%%%%%%%%%%%%%%%%%%%%%%% REFERENCES

%%%%%%%%%%%%%%%%%%%%%%%%%%%%%%%%%%%%%%%%%%%%%%%

%%%%%%%%%%%%%%%%%%%%%%%%%%%%%%%%%%%%%%%%%%%%%%%

%\newpage

%\eject

%\eject

%\end{multicols}
\end{document}